\documentclass[preprint]{aastex}
\usepackage{comment}
\usepackage{lineno}
\newcommand{\nhighlatf}{39}
\newcommand{\nhighlatbl}{58}
\newcommand{\nhighlatu}{163}
\newcommand{\nhighlatag}{2}
\newcommand{\nsrc}{262}
\newcommand{\nrdg}{2}
\newcommand{\nhighlataa}{36}
\newcommand{\nhighlatab}{2}
\newcommand{\nhighlatac}{0}
\newcommand{\nhighlatad}{1}
\newcommand{\nhighlatba}{13}
\newcommand{\nhighlatbb}{13}
\newcommand{\nhighlatbc}{23}
\newcommand{\nhighlatbd}{9}
\newcommand{\nhighlatca}{71}
\newcommand{\nhighlatcb}{26}
\newcommand{\nhighlatcc}{23}
\newcommand{\nhighlatcd}{43}

\newcommand{\nhighlatfc}{28}
\newcommand{\nhighlatblc}{50}
\newcommand{\nhighlatuc}{141}
\newcommand{\nhighlatagc}{2}
\newcommand{\nsrcc}{221}
\newcommand{\nrdgc}{2}
\newcommand{\nhighlatcaa}{27}
\newcommand{\nhighlatcab}{1}
\newcommand{\nhighlatcac}{0}
\newcommand{\nhighlatcad}{0}
\newcommand{\nhighlatcba}{11}
\newcommand{\nhighlatcbb}{11}
\newcommand{\nhighlatcbc}{20}
\newcommand{\nhighlatcbd}{8}
\newcommand{\nhighlatcca}{62}
\newcommand{\nhighlatccb}{21}
\newcommand{\nhighlatccc}{20}
\newcommand{\nhighlatccd}{38}

\newcommand{\nlowlat}{23}
\newcommand{\nrdgl}{0}
\newcommand{\nlowlatf}{0}
\newcommand{\nlowlatbl}{1}
\newcommand{\nlowlatu}{22}
\newcommand{\nlowlatag}{0}
\newcommand{\nlowlataa}{0}
\newcommand{\nlowlatab}{0}
\newcommand{\nlowlatac}{0}
\newcommand{\nlowlatad}{0}
\newcommand{\nlowlatba}{1}
\newcommand{\nlowlatbb}{0}
\newcommand{\nlowlatbc}{0}
\newcommand{\nlowlatbd}{0}
\newcommand{\nlowlatca}{10}
\newcommand{\nlowlatcb}{2}
\newcommand{\nlowlatcc}{0}
\newcommand{\nlowlatcd}{10}

\begin{document}
\title{The Fourth Catalog of Active Galactic Nuclei Detected by the  {\em Fermi} Large Area Telescope - Data Release 2}
\author{B.~Lott$^1$, D.~Gasparrini$^{2,3}$, S.~Ciprini$^{2,3}$  \\ The {\em Fermi}-LAT collaboration  \\
\vspace{10pt}
{\sl \small $^1$ Universit\'e de Bordeaux, CNRS, CENBG-IN2P3, F-33170 Gradignan, France} 
\\ {\sl \small $^2$ Istituto Nazionale di Fisica Nucleare, Sezione di Roma ``Tor Vergata", I-00133 Roma, Italy}
\\{\sl \small$^3$ Space Science Data Center - Agenzia Spaziale Italiana, Via del Politecnico, snc, I-00133, Roma, Italy}
}
\begin{abstract}
An incremental version  (4LAC-DR2) of the fourth catalog of active galactic nuclei (AGNs) detected by the  {\em Fermi}-LAT is presented. This version is associated with the second release of the 4FGL general catalog (based on 10 years of data), where the spectral parameters, spectral energy
distributions, yearly light curves, and associations have been updated for all sources. The new reported  AGNs include two radio galaxies and 283 blazars. We briefly describe the properties of the new sample and outline changes affecting the previously published sample. \\

\noindent Note: Users of this incremental release are requested to cite the original 4LAC paper (Ajello M. et al., 2020, ApJ, 892, 105). 

\end{abstract}
\section{Introduction}
The initial 4FGL general catalog  \citep[][referred to as 4FGL-DR1]{4FGL}  of sources detected by the  {\em Fermi}-Large Area Telescope  (LAT) was based on 8 years of data. An incremental version of that catalog \citep[4FGL-DR2, ][]{4FGLDR2} includes 723 additional sources detected with 10 years of data. We present the corresponding update\footnote{The 4LAC-DR2 sample comprises AGNs located at $|$b$|>10^\circ$, in keeping with the 4LAC-DR1 defining criterion. AGNs lying at $|$b$|<10^\circ$ constitute the low-latitude sample.  The 4LAC-DR2 and low-latitude files are available at https://www.ssdc.asi.it/fermi4lac-DR2/table-4LAC-DR2-h.fits and https://www.ssdc.asi.it/fermi4lac-DR2/table-4LAC-DR2-l.fits respectively. } to the 4LAC catalog \citep[][ herein referred to as 4LAC-DR1]{4LAC} comprising the  285 new 4FGL-DR2 AGNs.  These AGNs  are all blazars except for two radio galaxies. We refer the reader to the 4FGL-DR2 document for details on the gamma-ray analysis.  In 4FGL-DR2, variability is inferred by means of yearly light curves running over the 10-year period.  The association and classification procedures  are similar to those used for the initial 4FGL-DR1. The main change concerns the use of an updated version of the Radio Fundamental Catalog\footnote{rfc\_2020a available at \url{http://astrogeo.org/rfc/}} including 8\% more entries  in seeking counterparts.    

\section{The new 4LAC-DR2 and low-latitude samples}
 The overall sample of new sources  includes 39 flat-spectrum radio quasars (FSRQs),  59 BL~Lac objects (BL~Lacs), 185 blazar candidates of unknown types (BCUs), and two radio galaxies, NGC 3078 and NGC 4261.  The 4LAC-DR2 comprises  262  ($|$b$|>10^\circ$) new AGNs, while the low-latitude sample includes 23 others (all  BCUs except for one BL~Lac). SED-based classification is missing for 67 sources. Table \ref{tab:census} gives the census of the new AGNs. The same defining criterion for the clean sample (i.e., sources with no analysis flags) as in 4LAC-DR1 has been used.   

The  median photon index  of the new 4LAC-DR2  FSRQs is slightly larger (2.63 vs  2.45) than that of the DR1 sample, indicating softer spectra. The median redshift is very similar to 4LAC-DR1 (1.19 vs. 1.12). PKS~2318-087 (4FGL J2320.8$-$0823) with z=3.164 has the highest redshift of new 4LAC-DR2 FSRQs.  Four DR1 FSRQs have higher redshifts (up to 4.31).  A total of 16 new 4LAC-DR2 FSRQs are found to be variable. 

 The 58 new  4LAC-DR2 BL~Lacs have a median photon index similar to the DR1 ones  (2.05 vs.  2.03). The median redshift of the 28 BL~Lacs with measured values is  0.27, which is  comparable to that of 4LAC-DR1 (0.34). The  maximum redshift is 0.848 for RX J1438.3+1204 (4FGL J1438.6+1205), while the maximum redshift of 4LAC-DR1 BL~Lacs is 2.83. A total of 37  BL~Lacs  are found to be variable. 

Out of the 163 new 4LAC-DR2 BCUs, only 5 have measured redshifts. The  median photon index is similar to the corresponding DR1 value (i.e., 2.27 and 2.23 respectively). Only 3  BCUs  are found to be variable. 

NGC 3078 is a nearby (d=35 Mpc) compact-core-dominated galaxy \citep{Wro84}. NGC 4261 (d=30 Mpc) is a LINER Fanaroff-Riley type-I radio galaxy, whose LAT detection was first reported in \cite{deM20}.
    
\section{Changes to 4LAC-DR1}
For completeness,  we reiterate here the changes to 4LAC-DR1 AGNs as outlined in  the 4FGL-DR2 document.       
About 200 counterpart names of DR1 sources have been changed. It was noted that  blazar names from very large surveys (like 2MASS or WISE) were used for some 4FGL associations while more common names from radio catalogs were available.  Moreover some names referred to sources that are  offset by up to a few arcminutes from the real counterpart. We have replaced the non-radio names with those of radio counterparts whenever possible. Note that the positions reported in the 4FGL-DR1 (\texttt{RA\_Counterpart}, \texttt{DEC\_Counterpart}) fields were correct.     

Changes in associations of 4LAC-DR1 sources are:
\begin{itemize}
\item Three sources  (PKS 0736$-$770, TXS 1530$-$131, PKS 1936$-$623) were mistakenly classified as FSRQs. They  have been reclassified as BCUs. 
\item Recent follow-up observations of 4FGL blazars \citep{Fup_Mas12, Fup_Pag14, Fup_Pen17,Fup_Pen19} have enabled the classification of 132 former BCUs into 114 BL~Lacs and 14 FSRQs. 
 \item The latest version of the Radio Fundamental Catalog has enabled the association of six previously unassociated sources with  blazar candidates.  These sources are 4FGL J0550.9+2552 (NVSS J055119+254909), 4FGL J0803.5+2046 (GB6 B0800+2046), 4FGL J1347.4+7309 (NVSS J134734+731812), 4FGL J1606.6+1324 (NVSS J160654+131934), 4FGL J1738.0+0236 (PKS 1735+026), and  4FGL J2249.9+0452 (WISEA J225007.35+045617.3).
\item The tentative association of 4FGL J0647.7$-$4418 with the HMB  RX J0648.0$-$4418 reported in  4FGL-DR1 has been replaced by the association with the BCU SUMSS J064744$-$441946 following the multiwavelength investigation of  \citet{Assoc20_Marti}.
\end{itemize}

A total of 135 additional 4LAC-DR1 sources have been found variable thanks to the extended yearly light curves. 

\begin{figure}
\centering
\resizebox{14cm}{!}{\rotatebox[]{0}{\includegraphics{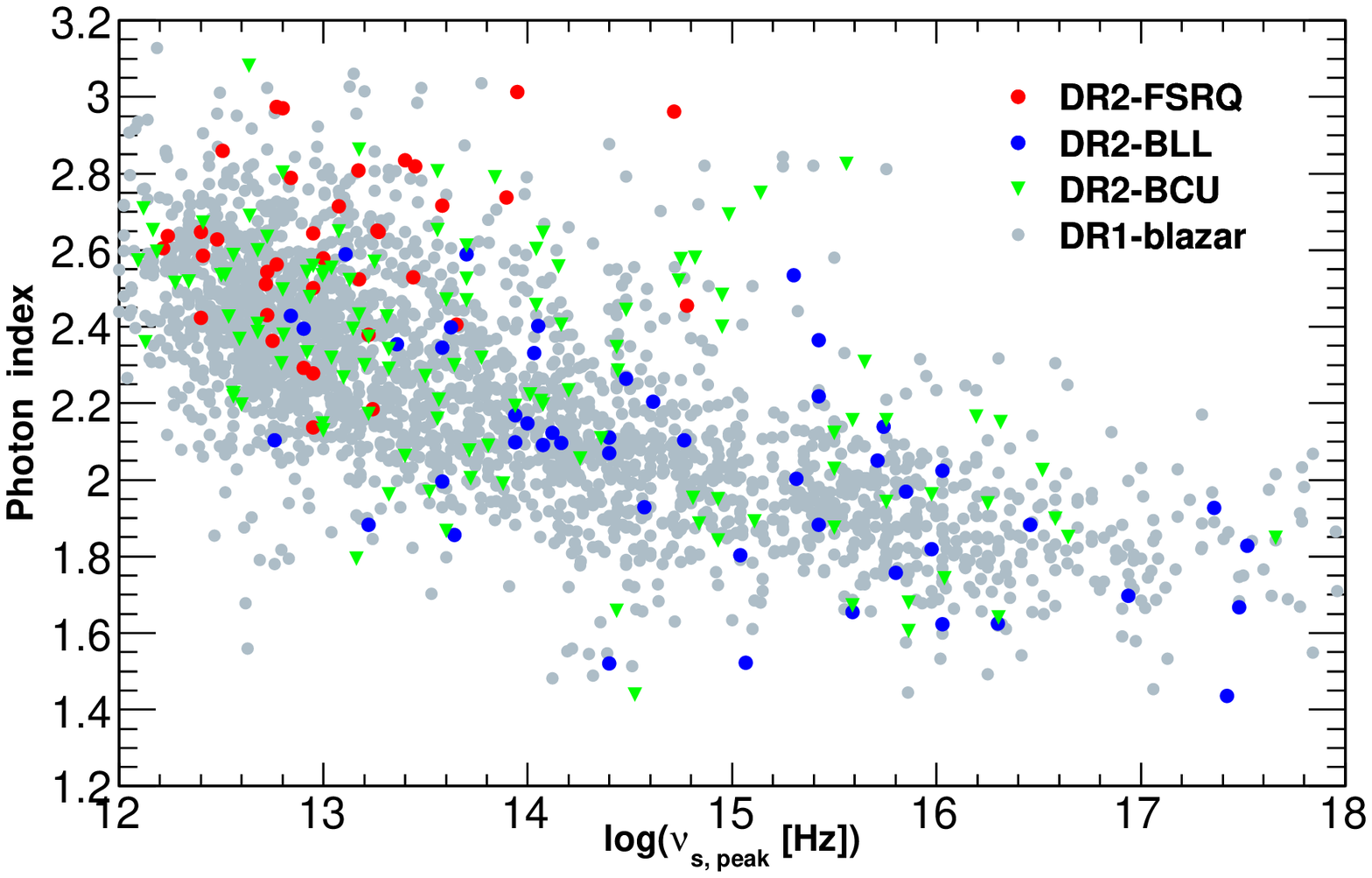}}}
\caption{Photon index vs. frequency of the synchrotron peak $\nu_\mathrm{s,peak}$ in the observer frame. Error bars have been omitted for clarity. }
\label{fig:index_nu_syn}
\end{figure}

\begin{deluxetable}{lrrr}
\tablecolumns{4}
\tabletypesize{\footnotesize}
\tablecaption{\label{tab:census}Census of new 4LAC-DR2 sources}
\tablewidth{0pt}
\tablehead{
\colhead{AGN type}&
\colhead{Entire 4LAC-DR2}&
\colhead{4LAC-DR2 Clean Sample\tablenotemark{a}}&
\colhead{Low-latitude sample}
}
\startdata
{\bf All}&{\bf \nsrc}&{\bf \nsrcc }&{\bf \nlowlat}\\
\\
{\bf FSRQ}&{\bf \nhighlatf}&{\bf \nhighlatfc}&{\bf \nlowlatf}\\
{\ldots}LSP& \nhighlataa & \nhighlatcaa & \nlowlataa\\
{\ldots}ISP& \nhighlatab &\nhighlatcab & \nlowlatab\\
{\ldots}HSP& \nhighlatac &\nhighlatcac &\nlowlatac\\
{\ldots}no SED classification & \nhighlatad &\nhighlatcad & \nlowlatad\\
\\
{\bf BL~Lac}&{\bf \nhighlatbl}&{\bf \nhighlatblc}&{\bf \nlowlatbl}\\
{\ldots}LSP& \nhighlatba & \nhighlatcba &\nlowlatba\\
{\ldots}ISP& \nhighlatbb &\nhighlatcbb &\nlowlatbb\\
{\ldots}HSP& \nhighlatbc &\nhighlatcbc & \nlowlatbc\\
{\ldots}no SED classification & \nhighlatbd & \nhighlatcbd & \nlowlatbd\\
\\
{\bf Blazar of Unknown Type}&{\bf \nhighlatu}&{\bf \nhighlatuc}&{\bf \nlowlatu}\\
{\ldots}LSP&  \nhighlatca& \nhighlatcca & \nlowlatca\\
{\ldots}ISP& \nhighlatcb &\nhighlatccb & \nlowlatcb\\
{\ldots}HSP& \nhighlatcc &\nhighlatccc & \nlowlatcc\\
{\ldots}no SED classification & \nhighlatcd &\nhighlatccd & \nlowlatcd\\
\\
{\bf Non-blazar AGN}&{\bf \nhighlatag}&{\bf \nhighlatagc}&{\bf \nlowlatag}\\
{\ldots}RG & \nrdg & \nrdgc & \nrdgl\\

\\
\enddata
\tablenotetext{a}{Sources with single counterparts and without analysis flags.}
\end{deluxetable}

\section{Acknowledgments}

\acknowledgments The \textit{Fermi} LAT Collaboration acknowledges generous
ongoing support from a number of agencies and institutes that have supported
both the development and the operation of the LAT as well as scientific data
analysis.  These include the National Aeronautics and Space Administration and
the Department of Energy in the United States, the Commissariat \`a l'Energie
Atomique and the Centre National de la Recherche Scientifique / Institut
National de Physique Nucl\'eaire et de Physique des Particules in France, the
Agenzia Spaziale Italiana and the Istituto Nazionale di Fisica Nucleare in
Italy, the Ministry of Education, Culture, Sports, Science and Technology
(MEXT), High Energy Accelerator Research Organization (KEK) and Japan
Aerospace Exploration Agency (JAXA) in Japan, and the K.~A.~Wallenberg
Foundation, the Swedish Research Council and the Swedish National Space Board
in Sweden. Additional support for science analysis during the operations phase is gratefully acknowledged from the Istituto Nazionale di Astrofisica in Italy and the Centre National d'\'Etudes Spatiales in France.
This work performed in part under DOE Contract DE-AC02-76SF00515.

\bibliography{Bibtex_4FGL_v1.bib}
\end{document}